\documentclass[
    aps,
    prx,
    reprint,
    superscriptaddress,
    longbibliography,
    floatfix,
]{revtex4-2}

\newcommand{\bra}[1]{\left \langle #1 \right \rvert}
\newcommand{\ket}[1]{\left \lvert #1 \right \rangle }

\usepackage{amsmath}
\usepackage{amstext}
\usepackage{amssymb}
\usepackage{graphicx}
\usepackage{import}
\usepackage{appendix}
\usepackage{lipsum}
\usepackage[pdfpagemode=UseNone,pdfstartview=FitH,colorlinks=true,linkcolor=blue,citecolor=blue,urlcolor=blue]{hyperref}
\usepackage[all]{hypcap}
\usepackage{subfigure}
\usepackage{siunitx}
\usepackage{cleveref}
\usepackage{float}
\crefname{equation}{Eq.}{Eqs.}
\Crefname{equation}{Eq.}{Eqs.}
\crefname{figure}{Fig.}{Figs.}
\Crefname{figure}{Fig.}{Figs.}
\crefname{section}{Sec.}{Secs.}
\Crefname{section}{Sec.}{Secs.}
\crefname{table}{Table}{Tables}
\Crefname{table}{Table}{Tables}
\crefname{appendix}{Appendix}{Appendices}
\Crefname{appendix}{Appendix}{Appendices}
\AddToHook{cmd/appendix/before}{\crefalias{section}{appendix}}  
\usepackage{tabularray}

\begin{document}

\title{Measurement-Induced State Transitions in Inductively-Shunted Transmons}

\newcommand{\Google}{Google Quantum AI, Santa Barbara, CA 93111, USA}

\author{Nicholas Zobrist}
\email{nzobrist@google.com}
\author{John Mark Kreikebaum}
\author{Mostafa Khezri}
\author{Sergei V.\ Isakov}
\author{Brian J.\ Lester}
\author{Yaxing Zhang}
\author{Agustin Di~Paolo}
\author{Daniel Sank}
\author{W.\ Clarke Smith}
\email{smithclarke@google.com}
\affiliation{\Google}
\begin{abstract}
Fast and high-fidelity qubit measurement plays a key role in quantum error correction.
In superconducting qubits, measurement is typically performed using a resonant microwave drive on a readout resonator dispersively coupled to the qubit.
Shorter measurement times require larger numbers of photons populating the readout resonator, which ultimately leads to undesired measurement-induced state transitions (MIST) of the qubit.
MIST can be particularly problematic because these transitions often leave the qubit in a high energy state, and for transmon qubits the MIST locations in readout parameter space drift as a function of qubit offset charge.
In transmon qubits, these drifts have been avoided using very large qubit-resonator detunings or dedicated offset charge biases.
In this work, we take an alternative approach and add an inductive shunt to the transmon to eliminate the offset charge dependence and stabilize the MIST.
We experimentally characterize MIST in several different inductively-shunted transmons, in agreement with quantum and renormalized semiclassical models for MIST.
These results extend to other inductively-shunted qubits.
\end{abstract}

\maketitle

\section{Introduction}\label{sec:intro}

Building a useful quantum computer out of error-prone physical qubits requires quantum error correction (QEC), where error syndromes are extracted using high-fidelity qubit measurements. In superconducting qubits, QEC has led to demonstrations of logical quantum memories outperforming their physical qubit constituents in several error correction codes~\cite{Ofek2016, Sivak2023, Acharya2025, He2025}. Dispersive coupling to a readout resonator~\cite{Blais2004} enables qubit measurement in these systems, reaching $\SI{0.25}{\percent}$ error within $\SI{100}{\nano\second}$~\cite{Swiadek2023,Spring2025}.

Improving the speed and fidelity of dispersive readout generally requires stronger microwave drives to populate the resonator with larger photon numbers. Unfortunately, more photons leads to unwanted resonances in the joint qubit-resonator system, also known as measurement-induced state transitions (MIST)~\cite{Sank2016}. In particular, qubit transitions to non-computational states due to MIST are difficult to revert and can have devastating consequences for QEC~\cite{Miao2023, Khezri2023}. Moreover, for transmon qubits, the specific locations in readout parameter space at which MIST occurs are highly dependent on the offset charge and therefore fluctuate significantly in time, degrading QEC performance~\cite{Khezri2023, Fechant2025, Malekakhlagh2022}. This instability can be mitigated using very large qubit-resonator detunings to remove MIST entirely~\cite{Kurilovich2025} or by stabilizing the offset charge~\cite{Fechant2025}.
An alternative strategy, which we pursue here, is to add an inductive shunt to the transmon. This provides a dc short across the superconducting islands, eliminating the dependence on static charge offsets and potentially stabilizing MIST~\cite{Verney2019, Smith2022, Xiao2025}.

Here we focus our attention on the inductively-shunted transmon (IST) qubit~\cite{Hassani2023, Kalacheva2024} which stands in the middle ground between the transmon~\cite{Koch2007} and the fluxonium~\cite{Manucharyan2009}. In particular, we explore the parameter regime where the charging energy is much smaller than the inductive energy and the Josephson energy, which are comparable. Like the transmon, the IST has a large capacitance that eases integration into large qubit arrays and a single-well potential that permits analytic models. Like the fluxonium, there is no dc offset-charge dependence due to the inductive shunt and the anharmonicity tunes over a wide range. These features have also made IST qubits a compelling platform for improved baseband CZ gates~\cite{Yuan2025} as well as longitudinal readout~\cite{Richer2017}. While MIST is now well understood in transmon qubits~\cite{Shillito2022, Dumas2024, Wang2025, Xia2025, Dai2026}, it has not been fully resolved for inductively-shunted qubits. In fluxonium qubits, MIST experiments at selected flux points have been compared to full quantum simulations~\cite{Bista2025}, metrics have been derived to reduce the need for lengthy simulations~\cite{Nesterov2024}, and effects introduced by Josephson junction array modes have been considered~\cite{Singh2025}.

In this paper, we characterize MIST in ISTs over a broad flux range, showing both agreement with full quantum simulations and stability in time, as expected from the insensitivity to dc offset charge. Furthermore, we introduce an effective Kerr model for the IST as well as a renormalized semiclassical model for MIST that significantly speed up processor design and calibration. We present data on two types of ISTs designed to operate at either the upper or lower flux insensitive points. In \cref{sec:characterization} we describe the fabricated device, an effective Kerr model, and the measurements used to determine the device parameters. In \cref{sec:mist} we present the MIST data and compare it to the renormalized semiclassical model and a full quantum simulation. We then demonstrate MIST stability and qualitatively compare it to data from a transmon qubit in a Willow processor~\cite{Acharya2025}. Finally, conclusions are drawn in \cref{sec:conclusions}.

\begin{figure}[ht]
\includegraphics{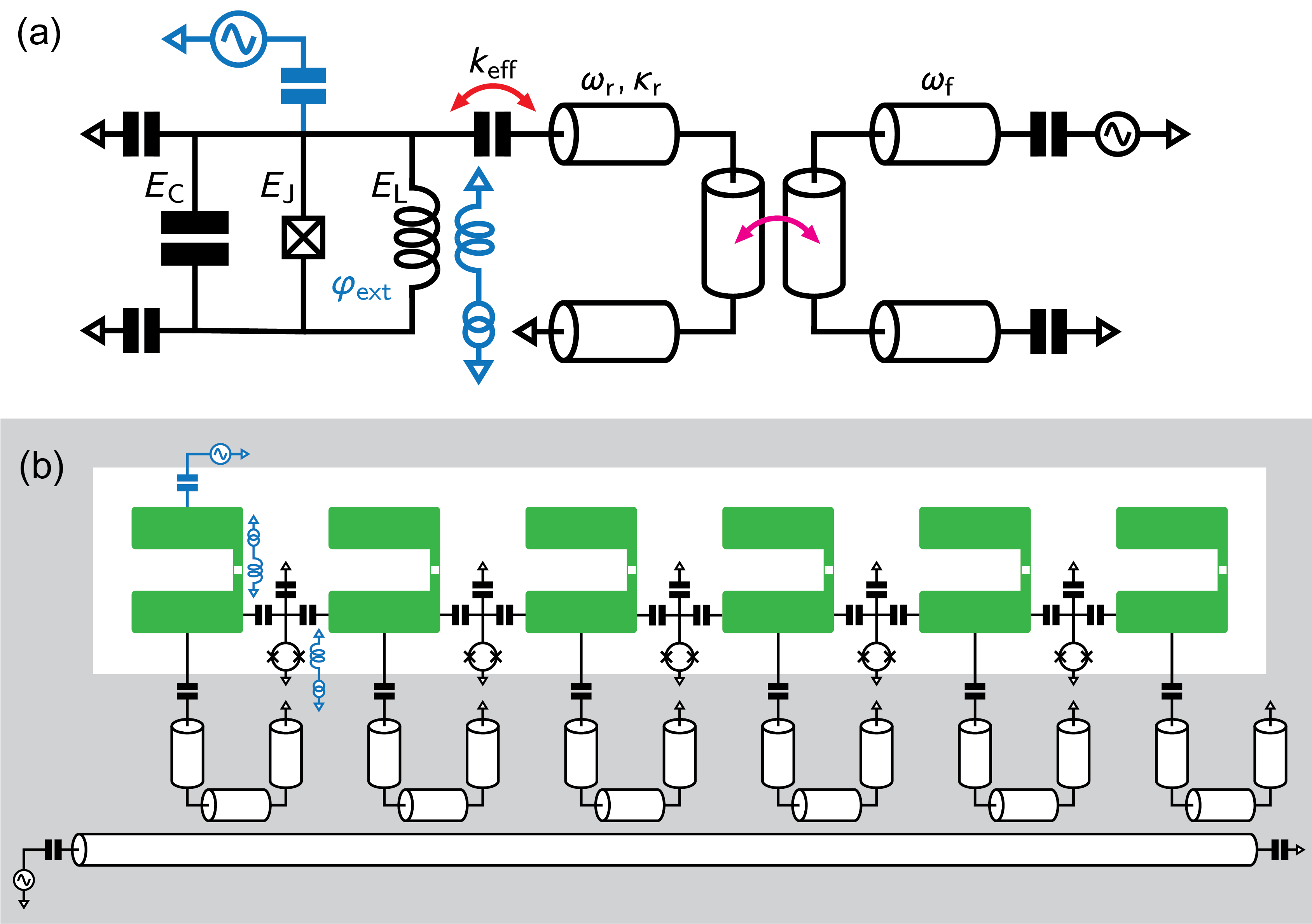}
\caption{Illustration of the device. (a) Circuit diagram of the qubit-resonator system without readout multiplexing. Each qubit has a separate capacitive drive and flux bias (blue). The readout resonator is coupled to a shared Purcell filter (pink). (b) The circuit schematic for a readout line. The qubit (green) and ground plane (gray) patterns are shown with equivalent circuit diagrams replacing the patterns for the couplers and readout circuit. Note the large capacitor pads of the IST, which ease connectivity in devices with large qubit numbers.
\label{fig:device}}
\end{figure}

\section{Qubit Characterization}\label{sec:characterization}

\subsection{Device Description}\label{subsec:device_description}
The nature of MIST depends on whether the qubit frequency $\omega_\mathrm{q}$ is above or below the readout resonator frequency $\omega_\mathrm{r}$~\cite{Sank2016, Khezri2023}. To access both cases, we designed a device with two groups of six flux tunable qubits. The first group was designed so that readout works well when $\omega_\mathrm{q} > \omega_\mathrm{r}$. Whereas readout works well in the second group when $\omega_\mathrm{q} < \omega_\mathrm{r}$. The primary consideration for both groups' design was the choice of a qubit resonator coupling that yields a $\chi$ comparable to $\kappa$ at the intended flux insensitive point. Each group is read out through a shared Purcell filter and readout line, which is shown in \cref{fig:device}\,(b). 

The qubits and resonators are fabricated on separate chips, and those chips are bump bonded together. Capacitive coupling between the qubits and resonators is achieved through the gap between the chips. The qubit fabrication, materials, and chip dimensions follow the gap-engineered qubits from Ref.~\cite{McEwen2024} with minor changes to resist thicknesses, evaporation angles, and oxidation parameters in order to introduce junction arrays for the IST's shunt inductance. The shunt inductance for each qubit is formed by a linear array of 9 Josephson junctions, each roughly 9 times larger in area than the small junction. All Josephson junctions were fabricated in a single lithography step with critical current density $J_c = \SI{22.5}{\ampere\per\centi\meter\squared}$, corresponding to a junction plasma frequency of $\SI{18.6}{\giga\hertz}$. We estimate the phase-slip rates for a single array junction in the first and second group to be $\SI{0.2}{\hertz}$ and $\SI{0.5}{\milli\hertz}$, respectively.
 
\subsection{Hamiltonian description and analytics}\label{subsec:hamiltonian_description}

Within a model where the junction array is treated as a linear inductance, we specify the full-circuit model of the qubit-resonator system by
\begin{equation}
    \begin{split}
    \hat H &= 4 E_\mathrm{C}\hat q^2 -E_\mathrm{J}\cos(\hat\phi-\varphi_\mathrm{ext}) + \frac{E_\mathrm{L}}{2}\hat \phi^2 + \hbar \omega_\mathrm{r}\hat a^\dagger \hat a \\
    &+2k_\mathrm{eff}\sqrt{\hbar\omega_\mathrm{r} E_\mathrm{C}}\,\hat q \,(\hat a + \hat a^\dagger),
    \end{split}
    \label{eq:full_circuit_model}
\end{equation}
where~$E_\mathrm{C}$, $E_\mathrm{J}$, and~$E_\mathrm{L}$ are the capacitive, Josephson, and inductive energies associated with the qubit. $\varphi_\mathrm{ext} = 2\pi \Phi_\mathrm{ext}/\Phi_0$ is the phase associated with the external flux $\Phi_\mathrm{ext}$, where~$\Phi_0=h/2e$ the superconducting flux quantum. $\omega_\mathrm{r}$ is the readout-resonator frequency. $k_\mathrm{eff}$ is the qubit-resonator coupling efficiency and is determined only by the capacitive network of the system. Additionally, $\hat\phi$ and~$\hat q$ are the qubit phase and Cooper pair number operators obeying $[\hat\phi, \hat q]=i$, while~$\hat a$ and~$\hat a^\dagger$ are bosonic ladder operators for the resonator satisfying~$[\hat a,\hat a^\dagger]=1$. 

We use the complete model in \cref{eq:full_circuit_model} to fit the measured spectrum in \cref{subsec:characterization} and to simulate qubit-readout dynamics in \cref{sec:mist}. This complete model can be computationally expensive, so we derive a simplified effective model of the qubit in~\cref{app:effective_hamiltonian} to analytically predict its frequency, anharmonicity, and the qubit-resonator dispersive shift. The result of the derivation is the effective model
\begin{equation}
    \hat{H}_\mathrm{q}/\hbar \approx \omega_\mathrm{q}\hat b^\dagger \hat b -\frac{\eta_\mathrm{q}}{2}\hat b^{\dagger 2}\hat b^2,
    \label{eq:kerr_model_simp}
\end{equation}
where~$\hat b$ and $\hat b^\dagger$ are bosonic ladder operators for the qubit satisfying~$[\hat b, \hat b^\dagger]=1$. $\omega_\mathrm{q}$ and $\eta_\mathrm{q}$ are dressed frequency and anharmonicity parameters linked to the circuit parameters via the effective IST-mode impedance. To analytically predict the qubit-resonator spectrum, we eliminate the transversal qubit-resonator coupling to second order in~$k_\mathrm{eff}$, capturing the dispersive and Lamb shifts. See Appendix~\ref{app:effective_hamiltonian} for details.

Because this model can be evaluated efficiently, it can be used during device calibration to predict transition frequencies at different flux biases. Additionally, we use it, here, to find an initial guess for the full numerical fit of the spectroscopy data described in the next section and plotted in \cref{fig:spectrum}. In particular, this effective model shows that the lowest few eigenenergies of the IST behave similarly to that of a transmon near the upper flux insensitive point.

\subsection{Characterization Measurements}\label{subsec:characterization}
The two groups of qubits operate near the upper and lower flux insensitive point respectively. After calibrating a readout which discriminates states $\left|0\right>$, $\left|1\right>$, and $\left|2\right>$, we perform several spectroscopy experiments to estimate the device parameters introduced in \cref{eq:full_circuit_model}. Between every experimental trial, each qubit is reset by flux biasing it to be on resonance with its readout resonator for 2 $\mu$s.

To measure $\omega_{10}$ versus flux bias, we dynamically bias the qubit from idle to a ``spectroscopy bias'', wait 2 $\mu$s to allow for the qubit frequency to settle, apply a 300 ns long variable frequency spectroscopy tone, and then finally return to the idle bias for state measurement. The frequency of the spectroscopy tone resulting in the highest $\left|1\right>$ population is then recorded for each spectroscopy bias as $\omega_{10}$. To further constrain our model, we measure $\omega_{21}$ near the spectroscopy bias by applying the same procedure with a $\left|0\right>$ to $\left|1\right>$ pulse after the 2 $\mu$s wait and before the spectroscopy pulse. We then record $\omega_{21}$ as the spectroscopy frequency resulting in the highest $\left|2\right>$ population for each spectroscopy bias. Finally, we find the maximum of the phase gradient of the readout tone across the resonator for each spectroscopy bias to estimate $\omega_\mathrm{r}$ versus qubit bias. 

While the highest $T_1$ values in these qubits are ${\sim}\SI{100}{\micro\second}$ (see \cref{app:coherence}), each qubit has a large number of strongly coupled resonant defects, which makes running experiments difficult. As such, we present data from one qubit from each group (Q1 \& Q2) with the least number of defects. During any data acquisition, only one qubit is measured at a time and the unused qubits are biased to the furthest possible frequency from the qubit being investigated. Similarly, the qubit couplers, which are not used for this experiment, are biased to their upper flux insensitive point near $\SI{10.5}{\giga\hertz}$ and $\SI{7}{\giga\hertz}$ respectively.

\begin{figure}[t!]
\centering
\includegraphics[width=\columnwidth]{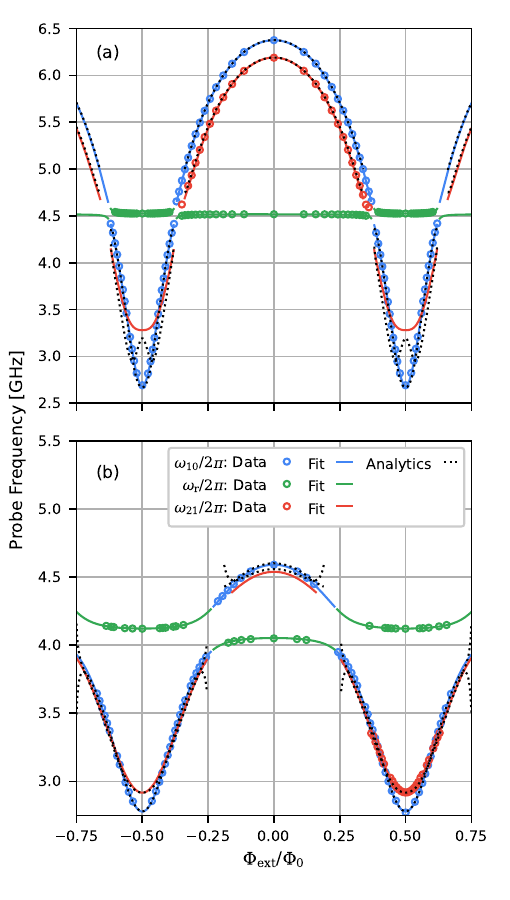}
    \caption{
    The spectrum of the qubit-resonator system vs. external bias for Q1 (a) and Q2 (b).
    Experimentally measured values are shown as circles.
    Fits to~\cref{eq:full_circuit_model} are shown as lines and the resulting full-circuit parameters are given in~\cref{tab:parameters}. Divergences near avoided level crossings are excluded from the plot for clarity. The black dotted lines show the qubit transitions as modeled by~\cref{eq:freq_and_eta}, including second-order shifts due to the coupling to the resonator. The analytic expressions are in excellent agreement with the data for the majority of fluxes. Deviations are observed around the qubit-resonator avoided level crossings, as expected from second-order perturbation theory, and near the lower flux insensitive point for Q1, where the IST potential approaches the onset of a double-well structure where the Kerr approximation fails. 
    \label{fig:spectrum}}
\end{figure}

The spectroscopy data and fit to the full model from \cref{subsec:hamiltonian_description} are plotted in \cref{fig:spectrum} along with the corresponding Kerr model from \cref{eq:kerr_model_simp}. The theory and experimental data are in excellent agreement over the majority of the flux tuning range, and the parameters extracted from the fits to \cref{eq:full_circuit_model} for the Q1 and Q2 are given in \cref{tab:parameters}. In this table we also include resonator linewidth, $\kappa_\mathrm{r}$, measurements determined from the ac Stark shift experiment described in Ref.~\cite{Sank2025}.

\begin{table}[h!]
\centering

\begin{tblr}{
  width = 0.8 \columnwidth,        
  colspec = {r | X[c] X[c]},   
  row{1} = {font=\bfseries}, 
  hline{1,Z} = {1.5pt},        
  hline{2} = {1pt},        
  hline{7} = {0.5pt},        
}    

    Parameter & Q1 & Q2 \\ 
    $E_\mathrm{C} / h$ [GHz]         & 0.398    & 0.198   \\
    $E_\mathrm{J} / h$ [GHz]         & 6.374     & 4.519     \\
    $E_\mathrm{L} / h$ [GHz]         & 7.142     & 8.859    \\ 
    $\omega_\mathrm{r} / 2 \pi$ [GHz]    & 4.523     & 4.111    \\
    $k_{\mathrm{eff}}$ [\%]     & 3.568     & 7.590      \\
    $\kappa_\mathrm{r} / 2 \pi$ [MHz]    & 3.639     & 6.376     \\

\end{tblr}
\caption{Summary of qubit and resonator parameters for the two qubits studied in this paper. The top set of parameters are extracted by fitting the spectrum shown in \cref{fig:spectrum} to \cref{eq:full_circuit_model}. The resonator linewidth, $\kappa_\mathrm{r}$, is found from a Lorentzian fit to the ac Stark shift experiment described in Ref.~\cite{Sank2025}.}
\label{tab:parameters}
\end{table}
\section{Measurement induced state transitions}\label{sec:mist}

In this section we show experimental data characterizing MIST in our IST device. Using the fitted Hamiltonian parameters from \cref{tab:parameters}, we also provide theoretical models that explain the experimental data. Finally, we comment on the trade-off between accuracy/speed and model complexity. 
\subsection{Experiment}\label{subsec:mist_experiment}
To experimentally measure MIST, we perform an experiment similar to Ref.~\cite{Khezri2023}. First, the qubit is prepared in either $\ket{0}$ or $\ket{1}$, and then flux biased to a target frequency.
Next, the readout resonator is driven for a duration of ${\sim}10/\kappa_\mathrm{r}$ to populate it with photons and reach steady state.
At every qubit frequency, this drive is chosen to be on resonance with the dressed frequency of the resonator for each initial state.
The drive is then turned off and the system idles for ${\sim}10/\kappa_\mathrm{r}$ to allow the resonator photons to decay out.
Finally, the population of the qubit is measured by driving the resonator with a weak tone calibrated to identify leakage outside of the qubit's computational subspace without introducing additional MIST.
This procedure is then repeated for various target frequencies and resonator drive powers.
To compare results with the model, the resonator drive power is converted to average resonator photons using the ac Stark shift \cite{Khezri2023, Sank2025}.

\begin{figure*}[th]
\centering
\includegraphics[width=\textwidth]{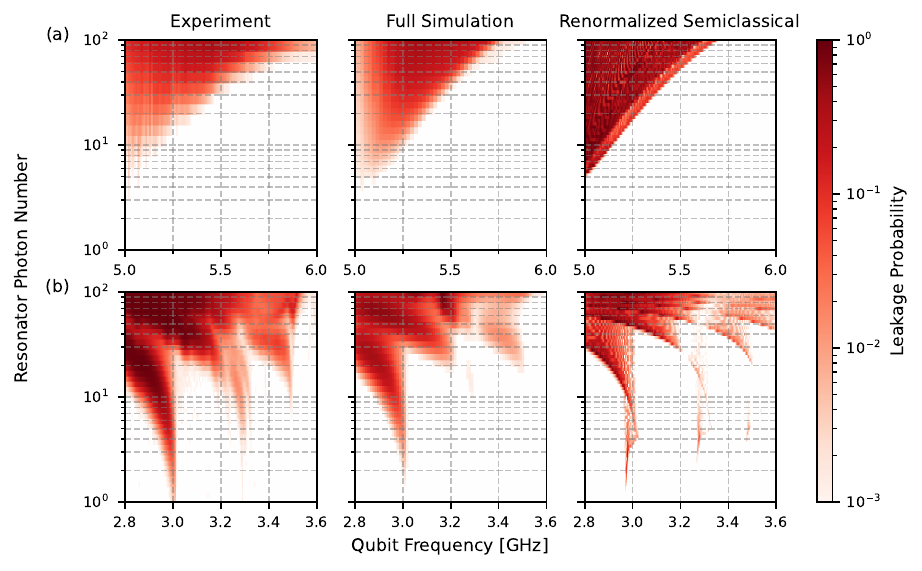}
    \caption{
    Ground state MIST experimental data and comparison to models. 
    Data and parameters are presented for two different designs of IST qubits Q1 (a) and Q2 (b).
    MIST behavior is significantly different when the qubit is above the readout resonator in frequency (a) or below it in frequency (b). Qubit frequency ranges were chosen to reveal the most MIST below 100 photons where the full simulation can be completed in a reasonable time.
    }\label{fig:mist}
\end{figure*}

\Cref{fig:mist} shows the experimental data for two different qubits, where the qubits are prepared in $\ket{0}$ (see \cref{app:experiments} for the case where the qubits are prepared in $\ket{1}$, which turned out to be qualitatively similar). Besides differing circuit parameters, the primary difference between the qubits is that Q1 in \cref{fig:mist}\,(a) is above its readout resonator in frequency, while Q2 in \cref{fig:mist}\,(b) is below its resonator in frequency, leading to significantly different MIST. We do not measure MIST on the opposite side of the resonator from the designed operation point in either case because qubit readout does not work well in this regime and flux tuning past the resonator removes significant amounts of leakage from the system. 

\subsection{Full Quantum Simulation}\label{subsec:mist_quantum}
To model MIST in ISTs, we first perform a full quantum simulation of the IST-resonator system. Simulation of this system, known as the Jaynes-Tavis-Cummings (JTC) model, is computationally expensive due to the large Hilbert space. Nevertheless, because it is the most accurate, we use a full simulation here to establish our understanding of the system and its behavior, and to benchmark semiclassical approximations.

We start with the qubit-resonator Hamiltonian of \cref{eq:full_circuit_model} and add a drive term to the resonator $\hbar\varepsilon(t)(\hat{a}e^{i\omega_\mathrm{d} t} + \hat{a}^\dagger e^{-i\omega_\mathrm{d} t})$.
Here $\varepsilon(t)$ is the drive amplitude and $\omega_\mathrm{d}$ is the drive frequency, chosen to be on resonance with the dressed resonator frequency.
We then simulate time evolution of the density matrix of the system using Lindblad master equation 
\begin{equation}\label{eq:master}
    \frac{\partial \hat{\rho}}{\partial t} = -i[H, \hat{\rho}] + \hbar\kappa_\mathrm{r} \left(\hat{a}\hat{\rho}\hat{a}^\dagger - \frac{1}{2}\{\hat{a}^\dagger\hat{a}, \hat{\rho}\}\right).
\end{equation}

For each simulation run, we initialize the system in the joint qubit-resonator eigenstate where qubit is either in $\ket{0}$ or $\ket{1}$, and the resonator is in its vacuum state.
The drive envelope, which is a square pulse, is then turned on for a duration of ${\sim}10/\kappa_\mathrm{r}$ to populate the resonator with $\bar{n}=\langle a^\dagger a \rangle$ steady state photons, and is turned off for a similar duration to allow the resonator to fully ring down.
We then record the final population of the qubit to characterize leakage outside of the initial state.
Here qubit and resonator circuit parameters are the same as those extracted via independent spectroscopy measurements and fits in \cref{tab:parameters}, except $\kappa_\mathrm{r}/2\pi=\SI{8}{\mega\hertz}$ for both qubits to speed up simulations.

To improve the efficiency of computations, for a given drive amplitude we chose the Hilbert space size of the resonator to be $\sim \bar{n} + 3\sqrt{\bar{n}}$ where $\bar{n}$ is the average resonator photon number, to avoid simulating unpopulated levels.
The results depend strongly on the number of IST levels (see discussion of \cref{subsec:mist_semiclassical}).
In the simulations where the IST is prepared in $\ket{0}$, we include 18 IST eigenstates; we found that including only 12 IST levels was insufficient to capture the experimental behavior in \cref{fig:mist}. Additionally, we need 24 IST levels to accurately model MIST when qubit starts in $\ket{1}$, and onset of MIST is above ${\sim}20$ photons (see \cref{app:experiments}).
Even with care not to make the Hilbert space too large, the MIST simulation for points with the most resonator photons takes ${\sim} 3.5$ hours per qubit frequency on a 90-core machine. 

The simulation results are shown in \cref{fig:mist}.
We observe good agreement with the experimental data for both qubit designs, where the model can predict the onset of MIST photons and the location of features in frequency.
We note that the agreement is not perfect. However, it captures the majority of experimental behavior.
The agreement between data and the model is sensitive to the exact qubit parameters chosen and can be improved with minor adjustments to the circuit parameters from the values extracted from the spectroscopy data fit (not shown here).

\subsection{Semiclassical Approximation}\label{subsec:mist_semiclassical}
A more computationally efficient approach to model MIST in superconducting qubits is to use a semiclassical approximation wherein the resonator is treated as a classical drive acting on the qubit, reducing the Hilbert space size to that of only the qubit.
MIST can then be modeled by calculating time dynamics of the qubit population under the effect of this drive \cite{Khezri2023}, or by identifying qubit-resonator avoided crossings in the semiclassical Floquet spectrum of the system \cite{Sank2016, Dumas2024}.
It has been shown that this approach faithfully explains experimental observations in transmon qubits \cite{Khezri2023}.

The usual semiclassical approximation relies on the resonator being occupied by many photons, and its validity degrades for small numbers of photons.
To see why, consider the coupling matrix element between qubit-resonator states $\ket{\text{qubit}, \text{resonator}}$ that have the same total number of excitations, i.e., $\ket{k, n} : k+n=\text{constant}$.
Under the rotating wave approximation (RWA), these states form an \emph{RWA strip}: a collection of states belonging to an excitation preserving subspace that interact with each other via RWA coupling terms, and do not interact with other RWA strips having a different total number of excitations \cite{Sank2016}.
Note that even though non-RWA terms that couple different RWA strips together are necessary to explain and model MIST, including them does not change the eigenspectrum of the qubit-resonator system significantly, and can be neglected as an initial approximation for the following discussion.

In the full quantum model of \cref{eq:full_circuit_model}, the coupling matrix element between RWA strip levels $\ket{k, n-k}$ and $\ket{k+1, n-k-1}$ having fixed total excitations of $n$ is $g_{k,k+1}\sqrt{n-k}$, where $g_{k, k+1}$ is the charge matrix element of the qubit, and $\sqrt{n-k}$ is the charge matrix element of the resonator.
When resonator photon numbers are sufficiently large, one can make the approximation $\sqrt{n-k}\approx\sqrt{n}$, similar to displacement transformation of a resonator \cite{Walls2008}.
This enables the resonator evolution to be treated independently of the qubit \cite{Khezri2016}, which then allow the quantum resonator to be replaced with a semiclassical drive $\sqrt{n-k}\rightarrow|\alpha|$ to get
\begin{align}\label{eq:semiclassical-simple}
    &\hat{H}_\text{int} =2k_\mathrm{eff}\sqrt{\hbar\omega_\mathrm{r} E_\mathrm{C}}\,\hat{q}\times2|\alpha(t)|\cos{\omega_\mathrm{r}t}, \\
    &\dot{\alpha} = -\frac{\kappa_\mathrm{r}}{2}\alpha -i\varepsilon(t). \label{eq:coherent}
\end{align}
\Cref{eq:coherent} describes the evolution of the coherent state $\alpha$ of a resonator that is driven on resonance, as in our experiments. 
For coherent states the average photon number is $\bar{n}=|\alpha|^2$ and its uncertainty (standard deviation) is $\sqrt{\bar{n}}$.
For large values of $\chi$ the resonator can become nonlinear, changing its resonance frequency and causing the drive to move off resonance at large photon numbers. The resonator nonlinearity is not accounted for here but could cause differences between the semiclassical and the full quantum models.

Levels which are not coupled in the full quantum model become coupled in the simple semiclassical model, significantly changing the eigenspectrum of the system.
In the quantum model, only levels up to $k < n$ can couple to each other with strength $g_{k,k+1}\sqrt{n-k}$, while higher qubit levels remain uncoupled.
For example for $n=3$ in $|k,n-k\rangle$, there is no coupling between RWA strip levels $|3, 3-3\rangle$ and $|4,3 -4\rangle$ because the later state does not exists in the quantum JTC model.
In the simple semiclassical model however, all neighboring qubit levels couple to each other with strength $g_{k, k+1}|\alpha|$.
If $|\alpha|^2=3$, $|3\rangle | \alpha\rangle$ couples to $|4\rangle |\alpha\rangle$ with strength $g_{3,4}\sqrt{3}$, as opposed to no coupling in the quantum model.
In other words, the energy structure and coupling terms of the quantum JTC model differ significantly between states near the low-energy sector and the bulk states in the high-energy sector.
The commonly used semiclassical model of qubit-resonator misses this difference.

The inclusion of couplings $k \geq n$ in the simple semiclassical model does not strongly affect MIST predictions for the transmon because the transmon has only a few levels confined to its potential well, and the charge matrix elements for levels outside of the well decrease exponentially with $k$, decoupling them from the rest.
However, the IST qubit levels are all confined, and charge matrix elements actually increase as ${\sim}\sqrt{k}$ for higher levels leading to different behavior.
This difference means that within the semiclassical model, all IST levels couple strongly to each other, especially for small values of $n$, where in the complete quantum model they do not interact with each other. The extra coupling terms lead to incorrect predictions of the spectrum of IST-resonator system (see \Cref{app:semiclassical_model}), making it inaccurate for modeling MIST.

This deficiency in the semiclassical model is improved by renormalizing its coupling matrix elements; we follow the idea introduced in Ref.~\cite{Khezri2023}, extending it to include non-RWA terms.
The interaction Hamiltonian for the renormalized semiclassical model is
\begin{equation}\label{eq:semiclassical-modified}
    \begin{split}
        \hat{H}_\text{int} =&\, 4k_\mathrm{eff}\sqrt{\hbar\omega_\mathrm{r} E_\mathrm{C}}\,\cos(\omega_\mathrm{r}t)\times \\
        &\sum_{k=0}^{M}\sum_{l=1}^{M-k} \text{Re}\left(\sqrt{|\alpha|^2-k}\right) g_{k,k+l}\ket{k}\bra{k+l} + \text{H.c.},
    \end{split}
\end{equation}
where $g_{k, k+l}$ are charge matrix elements of the IST eigenstates, and $M$ is the total number of levels included in the model.

With this renormalized semiclassical model, the interaction between levels is nonzero only when $k < |\alpha|^2$, resembling the JTC case discussed earlier.
Another benefit is that we no longer assume $\sqrt{n-k}\approx\sqrt{n}$ for small number of photons.
Note that this model is still only an approximation, and can not capture other subtleties present in the JTC model, e.g., for small photon numbers the coupling to levels outside of RWA strip (non excitation preserving terms) will be different at the bottom of the JTC energy structure compared to its bulk.

To model MIST using the renormalized semiclassical model, we first solve \cref{eq:coherent} to find $\alpha(t)$, where the drive amplitude $\varepsilon(t)$ is turned on and then off similar to the procedure described in  \cref{subsec:mist_quantum}.
We then numerically evolve the Schr\"{o}dinger equation time dynamics of \cref{eq:semiclassical-modified} for 30 IST eigenstates, and calculate the qubit populations at the end of the drive pulse.

Results are shown in \cref{fig:mist}.
The renormalized semiclassical model broadly agrees with the experimental data and reproduces its essential features.
However, there are some quantitative discrepancies beyond the $\pm\sqrt{\bar{n}}$ uncertainty of the model.
In particular, the model correctly predicts the frequencies of the major transitions but under predicts the onset photon number at which they appear. The agreement between data and model for ISTs is weaker than is typically seen for transmon qubits, but is not surprising given the challenges of approximating the IST energy structure as discussed above.
Importantly, without the renormalization modifications to the semiclassical model used here, the results would have been significantly different from experiments (see \cref{app:semiclassical_model}). We expect similar challenges in applying semiclassical models to other qubits with unbounded potentials such as fluxonium.

\begin{figure}[th]
\centering
\includegraphics[width=\columnwidth]{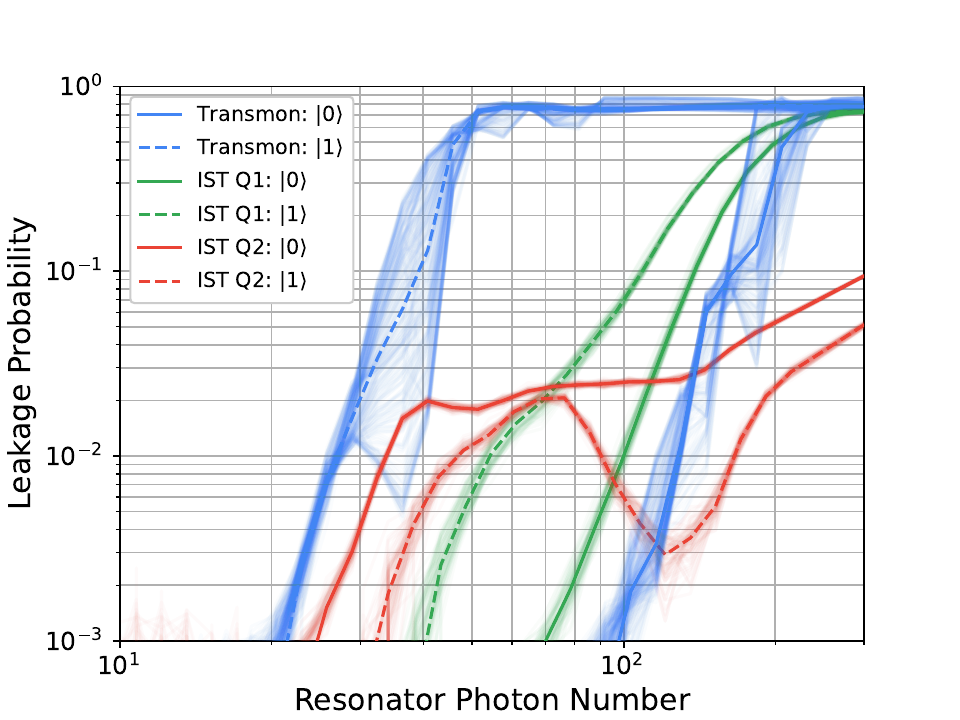}
    \caption{
    Leakage data for both IST and Transmon qubits vs. applied resonator photon number. The data was taken continuously over the course of 24 hours. The median of these datasets is shown as a solid color while the semi-transparent lines show individual runs. There is a qualitative difference in variability between the transmon and IST qubit types as expected from the lack of offset charge in IST qubits.}\label{fig:stability}
\end{figure}
\subsection{MIST Stability}\label{sec:mist_stability}

A central motivation for exploring ISTs is the expected lack of variability of MIST due to the insensitivity to slow-varying charge offsets. Note that the IST remains sensitive to higher frequency components of offset charge noise, which stems from the inductive shunt acting as a $\sim \SI{3}{\giga\hertz}$ high-pass filter with respect to charge fluctuations. To investigate this, we focus on a single qubit frequency and measure leakage versus resonator photon number for 24 hours. Each dataset takes about 4 minutes to complete. Q1 is biased to 6.19 GHz and Q2 to 2.77 GHz. Results are shown in \cref{fig:stability}.

Although any comparison of MIST between IST and transmon qubits must be qualitative, Willow qubits have similar resonator frequencies and qubit anharmoncities as Q1. We include leakage data for a Willow qubit in the same configuration as Q1 in \cref{fig:stability}. The variability in the transmon versus the IST leakage curves is distinct and shows that MIST in the IST is significantly more stable. Furthermore, this measurement shows that low-frequency offset charge fluctuations are the dominant source of MIST instability, rather than two-level systems, quasiparticles, ionizing radiation, junction array offset charges, etc.
\section{Conclusions}\label{sec:conclusions}

ISTs are promising candidates for quantum error correction. Their main advantages are a large anharmonicity at low frequency with no offset charge noise, large capacitance for coupling to neighboring qubits, and a stable MIST landscape that simplifies readout optimization.
To explore these properties, we fabricated two types of IST qubits with frequencies both above and below the readout resonator, demonstrating $T_1$ times up to 100 $\mu$s.
The measured spectra were well predicted by an approximate Kerr Hamiltonian, and the main features of MIST were captured by a renormalized semiclassical model without needing to include effects of resonances in the junction chain. Finally, we observed qualitatively more stable MIST features in the IST as compared to a similar transmon.

To accurately predict the observed MIST features, the quantum nature of photons in the resonator was taken into account. A typical semiclassical treatment would have represented the resonator as a drive, where removal of energy from the resonator does not change its state. This approximation works when modeling driven processes, like parametric gates, where the drive is strong (i.e. contains many photons) but weakly coupled. It also works in modeling MIST in the transmon because the transmon's charge matrix elements decrease rapidly for states above the edge of the cosine potential, so processes where the transmon would accept many resonator photons are suppressed.
In the IST, on the other hand, the unbounded potential energy leads to large charge matrix elements for all levels, so approximating the resonator as a drive fails. Additionally, because of the unbounded potential energy, MIST in ISTs cannot be described as an ionization process.

This work represents a step forward in the MIST modeling of inductively-shunted qubits where we have compared data to theory over a wide range of operation frequencies and flux bias conditions. We expect that these modeling techniques will allow for the design of highly performant readout circuits in these types of qubits that exceed the stability of that for the transmon.

\begin{acknowledgments}
We are grateful to the Google Quantum AI team for building, operating, and maintaining software and hardware infrastructure used in this work. We thank Jayameenakshi Venkatraman, Xu Xiao, and Michel Devoret for their essential contributions to the conception of the inductively-shunted transmon.
\end{acknowledgments}
\section*{Contributions}\label{sec:contributions}
D.S. and W.C.S. conceived of the experiment. 
N.Z., J.M.K., A.D.P., and W.C.S. determined the device parameters and layout. 
N.Z. and W.C.S constructed the experimental apparatus. 
N.Z. took MIST, spectroscopy, and readout characterization data. 
B.L. took coherence data. 
M.K. and S.V.I modeled, simulated, and analyzed MIST. 
Y.Z. derived noise properties based on magnetic coupling to the bias line for the device design. 
A.D.P. developed the circuit theory. 
J.M.K. fabricated the device.
All authors contributed to the writing of the paper.

\appendix
\section{Effective IST Hamiltonian}\label{app:effective_hamiltonian}

To derive \cref{eq:kerr_model_simp}, we start from \cref{eq:full_circuit_model} only including terms for the qubit. We introduce an impedance parameter, $\xi$ and a displacement parameter, $\phi^*$, such that $\hat \phi\to\hat \phi + \phi^*$. Following this transformation, we rewrite the qubit Hamiltonian in terms of ladder operators~$\hat b$ and~$\hat b^\dagger$ defined by
\begin{equation}
    \begin{split}
        \hat \phi &= \sqrt{\xi}(\hat b+\hat b^\dagger),\\
        \hat q &= -\frac{i}{2\sqrt{\xi}}(\hat b-\hat b^\dagger).
    \end{split}
    \label{eq:qubit ladder}
\end{equation}
Note that we employ a different convention for the qubit charge operator when expressed in terms of bosonic ladder operators compared to the one we use in~\cref{eq:full_circuit_model} for the readout resonator. The later is a more conventional choice in the context of readout, and these different conventions do not impact the results.
Expanding the cosine potential in a normal-ordered series~\cite{Petrescu2023} and setting to zero both linear and off-diagonal quadratic contributions, we arrive at the conditions
\begin{equation}
    \begin{split}
    -E_\mathrm{J}e^{-\xi/2}\sin\varphi_\mathrm{ext}^* + E_\mathrm{L}\phi^*&=0,\\
    (\xi/\xi_\mathrm{L})^2 + (\xi/\xi_\mathrm{J})^2e^{-\xi/2}\cos\varphi_\mathrm{ext}^* - 1&=0,
    \end{split}
    \label{eq:xi_equations}
\end{equation}
where~$\xi_\mathrm{J,L} = \sqrt{2 E_\mathrm{C}/E_\mathrm{J,L}}$. We numerically solve Eqs.~\eqref{eq:xi_equations} to determine the values of our displacement ($\phi^*$) and impedance ($\xi$) parameters. The remaining IST Hamiltonian reads
\begin{equation}
    \begin{split}
    \hat{H}_\mathrm{q}/\hbar &= \omega_0\hat b^\dagger \hat b -\frac{\eta_0}{2}\hat b^{\dagger 2}\hat b^2\\
    &-\frac{\eta_0}{3}(\hat b^{\dagger 3}\hat b + \hat b^{\dagger}\hat b^3)- \frac{\eta_0}{12}(\hat b^{\dagger 4} + \hat b^4)\\
    &+\eta_0\xi^{-1/2}\tan\varphi^*_\mathrm{ext}(\hat b^{\dagger 2}\hat b + \hat b^\dagger \hat b^2)\\
    &+\frac{\eta_0}{3}\xi^{-1/2}\tan\varphi^*_\mathrm{ext}(\hat b^{\dagger 3}+ \hat b^3)\\
    &+ \mathcal{O}(\xi),
    \end{split}
    \label{eq:kerr_model}
\end{equation}
where we have introduced~$\varphi_\mathrm{ext}^*=\varphi_\mathrm{ext}-\phi^*$, alongside the bare frequency and anharmonicity parameters
\begin{equation}
    \begin{split}
    \hbar\omega_0 &= 4E_\mathrm{C}\xi^{-1},\\ 
    \hbar\eta_0 &= E_\mathrm{C} - \frac{E_\mathrm{L}\xi^2}{2}.
    \end{split}
    \label{eq:kerr_model_bare_params}
\end{equation}
In a low-impedance limit where terms of order at least~$\xi$ can be neglected, \cref{eq:kerr_model} provides a form of the IST Hamiltonian similar to that of a transmon qubit--a weakly anharmonic oscillator (or Kerr) model. This approximation is valid for the devices considered in this work.

We leverage~\cref{eq:kerr_model} to derive analytical expressions for the qubit frequency, and anharmonicity. We do so by splitting the Kerr-model Hamiltonian in diagonal and off-diagonal components with respect to the Fock basis. Next, we treat the off-diagonal terms as a perturbation, and eliminate it via a second-order Schrieffer-Wolff transformation, arriving at the effective IST frequency and anharmonicity parameters
\begin{equation}
    \begin{split}
    \omega_\mathrm{q} =& \omega_0 -\frac{7 \eta_0^2}{24\left(\omega_0 - \frac{3\eta_0}{2}\right)}-\frac{5 \eta_0^2}{24\left(\omega_0 - \frac{5\eta_0}{2}\right)}\\
    &-\frac{16 \eta_0^2\tan^2\varphi_\mathrm{ext}^*}{9\xi(\omega_0 - \eta_0)}-\frac{8 \eta_0^2\tan^2\varphi_\mathrm{ext}^*}{9\xi(\omega_0 - 2\eta_0)},\\
    \eta_\mathrm{q} =& \eta_0 -\frac{5\eta_0^2}{8\left(\omega_0 - \frac{3\eta_0}{2}\right)}+\frac{5\eta_0^2}{8\left(\omega_0 - \frac{7\eta_0}{2}\right)}+\frac{9\eta_0^2}{4\left(\omega_0 - \frac{5\eta_0}{2}\right)}\\
    &-\frac{52\eta_0^2\tan^2\varphi_\mathrm{ext}^*}{9\xi(\omega_0 - \eta_0)}+\frac{20\eta_0^2\tan^2\varphi_\mathrm{ext}^*}{9\xi(\omega_0 - 3\eta_0)}+\frac{92\eta_0^2\tan^2\varphi_\mathrm{ext}^*}{9\xi(\omega_0 - 2\eta_0)}.
    \end{split}
    \label{eq:freq_and_eta}
\end{equation}
As a result of this transformation, \cref{eq:kerr_model} simplifies to the model provided in the main text in \cref{eq:kerr_model_simp}.

Finally, we reincorporate the resonator terms in \cref{eq:full_circuit_model} and treat the interaction with the qubit perturbatively, performing a second Schrieffer-Wolff transformation to eliminate the qubit-resonator coupling to second order in $k_\mathrm{eff}$. As a result, we arrive at an expression for the dispersive shift
\begin{equation}
    2\chi = 4g^2_\mathrm{qr}\left[\frac{\omega_\mathrm{q}-\eta_\mathrm{q}}{\omega_\mathrm{r}^2 - (\omega_\mathrm{q}-\eta_\mathrm{q})^2}-\frac{\omega_\mathrm{q}}{\omega_\mathrm{r}^2 - \omega_\mathrm{q}^2}\right],
    \label{eq:dispersive_shift}
\end{equation}
where
\begin{equation}
    \hbar g_\mathrm{qr} = k_\mathrm{eff}\sqrt{\hbar\omega_\mathrm{r}\,E_\mathrm{C}\xi^{-1}} = 
    \frac{\hbar}{2}k_\mathrm{eff}\sqrt{\omega_\mathrm{r}\omega_0}.
\end{equation}
\cref{eq:dispersive_shift} is in agreement with what is known for transmons, highlighting the usefulness of the Kerr model described in this section. In a similar way, we can compute the Lamb shifts of the qubit energy-levels in the presence of coupling to the resonator, which are important to correctly predict the spectra in \cref{fig:spectrum}.

\begin{figure*}[th]
\centering
\includegraphics[width=\textwidth]{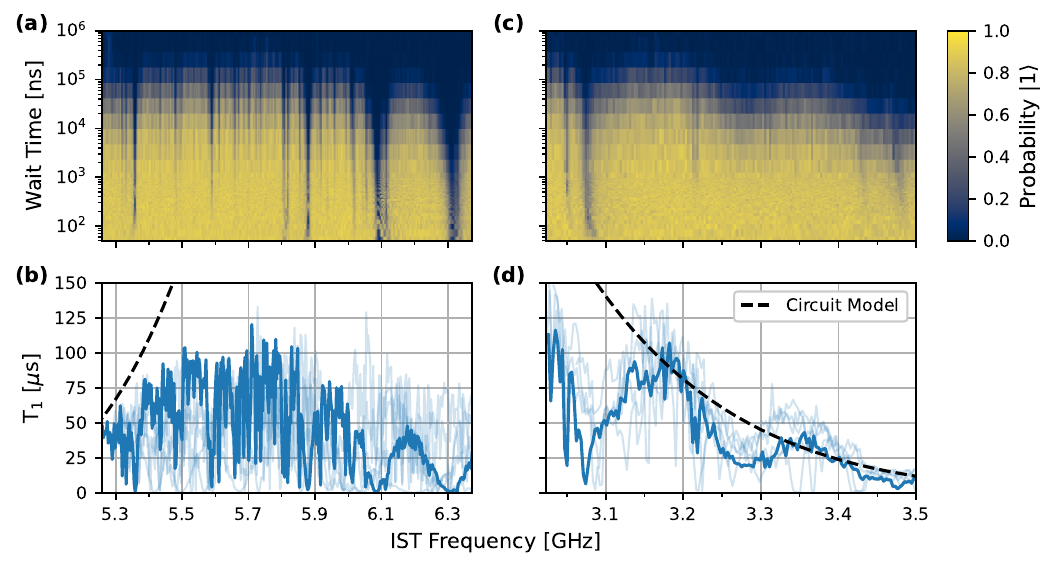}
    \caption{
    Coherence data for IST qubits Q1 (a, b) and Q2 (c, d). Panels (a) and (c) show the population versus time at different qubit frequencies. Pannels (b) and (d) report the fitted exponential decay time associated with each frequency. The dark curves are the data for Q1 and Q2. Other qubits on the chip are plotted for comparison in the lighter color. Note that at some of the $T_1$ minima, the population oscillates over time, making this fit value not meaningful. The dashed lines show the expected $T_1$ using the measured qubit parameters in \cref{tab:parameters} and the simulated readout circuit. 
    }\label{fig:t1}
\end{figure*}

\section{Qubit Coherence}\label{app:coherence}
The qubits in this experiment were significantly impacted by defects in their $T_1$ spectrum, shown in \cref{fig:t1}. All qubits have several strongly coupled defects, most of which coherently swap excitations with the qubit. These defects do not affect the MIST data in the main text, since it is for state $|0\rangle$, but could cause excess $|0\rangle$-state MIST in the $|1\rangle$ state data presented in \cref{app:experiments}. Additionally, the number of defects in some of the qubits makes characterization difficult which is why this paper focuses on Q1 and Q2.

We also model the Purcell decay through the readout system using Fermi's golden rule

\begin{equation} \label{eq:t1}
 T_1^{-1} = \frac{2 \omega_{10}}{\hbar}\text{Re}\left[Z_\text{in}(\omega_{10})\right] \big| \big\langle 0 \big| \hat{Q} \big| 1 \big\rangle \big|^2,
\end{equation}
where $\hat{Q} = 2e \, \hat{q}$ \cite{Schoelkopf2003}. The qubit parameters measured in \cref{tab:parameters} are used in \cref{eq:t1} to compute the matrix element. For the real part of the impedance seen by the qubit junction $\text{Re}\left[Z_\text{in}(\omega_{10})\right]$, we use the simulated circuit parameters extracted from a 3D electromagnetic solver including an internal resonator and filter loss of $Q_i=150,000$ which we have found to be a good estimate for the low power loss in resonators of similar design. The $T_1$ predicted by the model agrees well with the qubit coherence closest to the resonator where it is limited by the readout circuit. For Q2, we see oscillations in $T_1$ across all qubits. This effect likely comes from an impedance mismatch outside of our qubit chip not taken into account by the circuit model.
\section{Semiclassical model for IST}\label{app:semiclassical_model}

\begin{figure*}[th]
\centering
\includegraphics[width=0.9\textwidth]{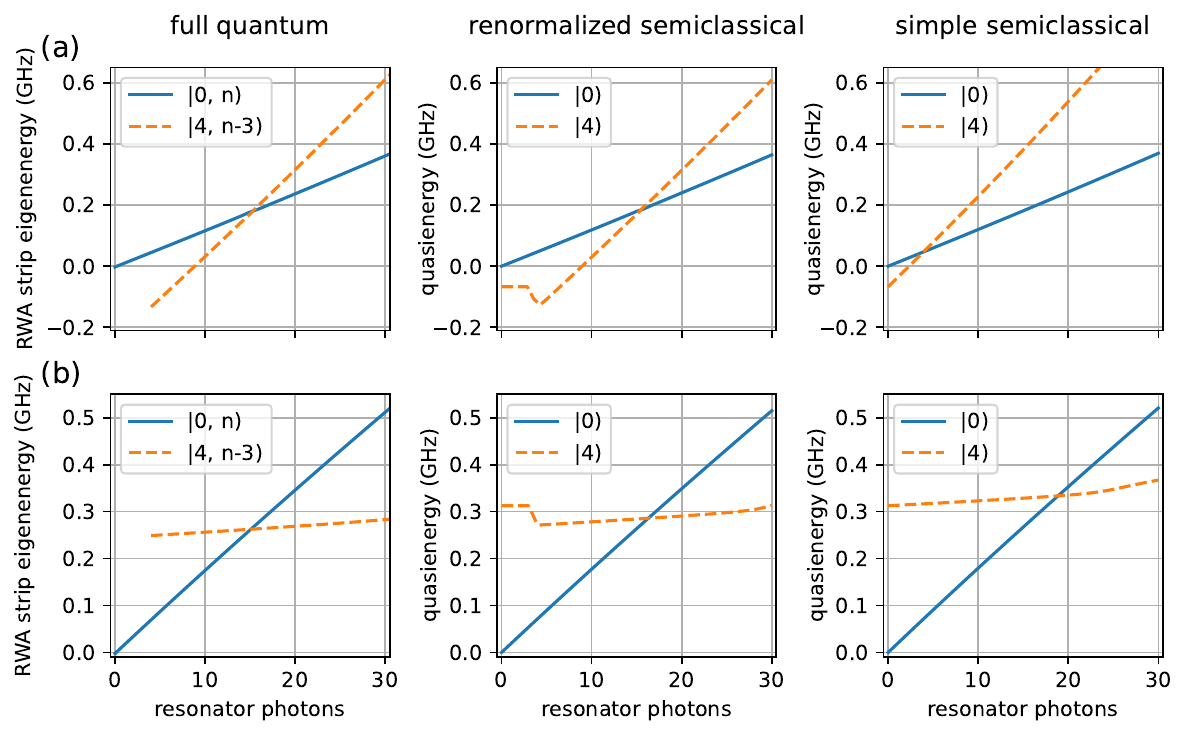}
    \caption{
    Comparison of qubit-resonator spectrum using three different models.
    (a) Spectrum for an IST qubit.
    IST semiclassical spectrum is very different compared to the quantum model if renormalization is not applied.
    (b) Spectrum for a transmon qubit.
    For transmon, renormalization of semiclassical model does not have a significant effect on the spectrum compared to the quantum model.
    Both qubits are modeled with 20 eigenlevels and use parameters relevant to typical experiments.
    }\label{fig:model_comparisons}
\end{figure*}

In this appendix we provide evidence that the renormalization of semiclassical approximation is \emph{necessary} to properly model IST qubits.
Additionally, we show that this renormalization does not lead to significant changes for the transmon qubit.

We start by numerically calculating the spectrum of the full quantum model of qubit-resonator system of \Cref{eq:full_circuit_model}, and look at RWA strip eigenenergies in this system.
These were introduced in \cite{Sank2016} as eigenenergy of excitation preserving levels minus the resonator frequency for the total excitation number, formally $E_{\overline{|k, n-k\rangle}} - n\hbar\omega_\mathrm{r}$.
MIST is caused by resonances between these eigenenergies, therefore, properly capturing the spectrum enables accurate MIST prediction.
Here we consider the full quantum model as the ground truth for other approximations.

Next, we numerically calculate the Floquet spectrum or quasienergies of a periodically driven Hamiltonian \cite{Dumas2024} for different semiclassical models that were presented in the main text.
We calculate these vs. fixed resonator photons $|\alpha|^2$ for the simple semiclassical model of \Cref{eq:semiclassical-simple}, and also for the renormalized model of \Cref{eq:semiclassical-modified}.

\Cref{fig:model_comparisons} shows the spectrum calculated via three different models, for an IST qubit (a) and a transmon qubit (b).
Here we use some representative parameters as an example, chosen such that for both qubits there is a crossing between $|0\rangle$ and $|4\rangle$ at $\sim 15$ photons, and we note that the following conclusions remain the same for other parameters.
It is clear from the figure that the quasienergies of the semiclassical model represent the same physical quantity as RWA strip eigenenergies of the quantum model.

Results of \Cref{fig:model_comparisons} shows that for the IST qubit, the renormalized semiclassical approximation can properly recreate the quantum crossing at $\sim 15$ photons, but the simple semiclassical model fails and predicts a crossing at $\sim 5$ photons.
For the transmon however, both renormalized and simple models predict crossings much closer to quantum value of $\sim 15$ photons. Although, the renormalized model still shows a better agreement.
Therefore, it is essential that renormalized semiclassical approximations be used for IST and other qubits with similar energy structure such as fluxonium, otherwise MIST predictions will be inaccurate.
\section{$|1\rangle$ MIST Experiment}\label{app:experiments}

\begin{figure*}[th]
\centering
\includegraphics[width=\textwidth]{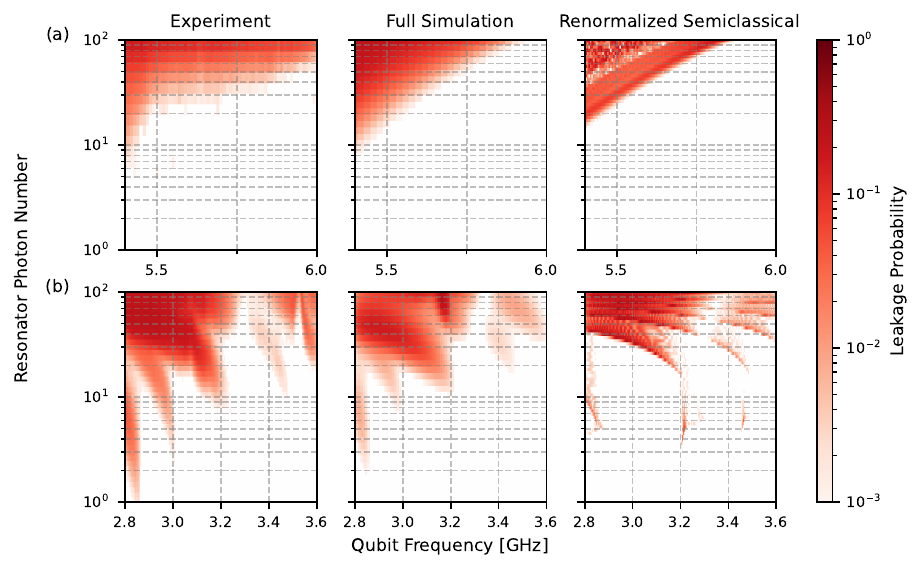}
    \caption{
    Excited state MIST experimental data and comparison to models. 
    Data is presented for two different designs of IST qubits, Q1 (a) and Q2 (b), which are the same qubits as the ones discussed in main text. Q1 data does not extend as low in frequency as for $|0\rangle$ in \cref{fig:mist} because the resonator has an avoided level crossing which prevents driving on resonance. 
    }\label{fig:mist_excited}
\end{figure*}

In this section we provide experimental data and comparison with models for when the qubit is initially prepared in the excited state.
The result is presented in \cref{fig:mist_excited}.
While the models qualitatively capture most of the MIST behavior, there are quantitative differences.

The excited state experimental data is difficult to interpret, in general, in the presence of resonant defects and Purcell loss which both induce qubit decay. If $T_1$ during the readout is on the order of the readout time, the qubit may decay to $|0\rangle$ and then MIST might occur. Some discrepancies are expected between data and model because of this effect. The missing feature at ${\sim}3$~GHz in the semiclassical model for \cref{fig:mist_excited}(b), for example, is due to $T_1$ decay from Purcell effect not being included in that model [note similar feature in \cref{fig:mist}(b)].
The full simulation catches this feature, but, while the model includes resonator loss, it does not predict its onset photon number accurately because the model does not contain an accurate description of the Purcell decay frequency dependence.

Additionally, for Q2, we compare leakage to $|3\rangle$ and above to the model because the experimental data showed significant $|2\rangle$ leakage irrespective of readout power at some flux biases. We attribute these effects to stray drives on the qubit from electronics and resonant defects discussed in \cref{app:coherence}. 

\clearpage

%

\end{document}